\documentclass[aps, preprint, superscriptaddress]{revtex4-1}
\usepackage[utf8]{inputenc}
\usepackage[T1]{fontenc}
\usepackage{amsmath}
\usepackage{amsfonts}
\usepackage{amssymb}
\usepackage{graphicx}
\usepackage{lineno}
\usepackage{xcolor}
\usepackage{hyperref}

\begin{document}

\title{Emergence of a Novel Phase in Population and Community Dynamics Due to Fat-Tailed Environmental Correlations}

\author{Immanuel Meyer}
\affiliation{Department of Physics, Bar-Ilan University, Ramat-Gan IL52900, Israel}
\author{Ami Taitelbaum}
\affiliation{Racah Institute of Physics, The Hebrew University, 9190401 Jerusalem, Israel}
\author{Michael Assaf}
\affiliation{Racah Institute of Physics, The Hebrew University, 9190401 Jerusalem, Israel}
\author{Nadav M. Shnerb}
\affiliation{Department of Physics, Bar-Ilan University, Ramat-Gan IL52900, Israel}

\begin{abstract}
\noindent 
Temporal environmental noise (EN) is a prevalent natural phenomenon that controls population and community dynamics, shaping the destiny of biological species and genetic types. Conventional theoretical models often depict EN as a Markovian process with an exponential distribution of correlation times, resulting in two distinct qualitative dynamical categories: quenched (pertaining to short demographic timescales) and annealed (pertaining to long timescales). However, numerous empirical studies demonstrate a fat-tailed decay of correlation times. Here, we study the consequences of power-law correlated EN on the dynamics of isolated and competing populations. We reveal the emergence of a novel intermediate phase that lies between the quenched and annealed regimes. Within this phase, dynamics are primarily driven by rare, yet not exceedingly rare, long periods of almost-steady environmental conditions. For an isolated population, the time to extinction in this phase exhibits a novel scaling with the abundance, and also a non-monotonic dependence on the spectral exponent. 
\end{abstract}
\maketitle

The dynamics of populations and genetic types is influenced by two key factors: the typical lifespan of an isolated population, which is governed by its intrinsic dynamics, and the population's ability to outcompete other types, such as the competition between wild-type and mutant populations. Understanding these factors and their dependence on system parameters like growth rate, competition strength, fitness, and population size, is a key aspect of theoretical research in evolution, population genetics and community ecology~\cite{karlin1981second,ewens2004mathematical}. Yet, the unpredictable and erratic nature of biological environments introduces considerable variability in these parameters, significantly impacting the outcomes of the dynamics. Consequently, substantial theoretical, empirical, and experimental efforts have been dedicated to exploring the influence of environmental stochasticity, or environmental noise (EN), in these scenarios~\cite{lande2003stochastic,drake2006extinction,blythe2007stochastic,mustonen2008molecular,assaf2013cooperation,cvijovic2015fate,hidalgo2017species,wienand2017,danino2018fixation,taitelbaum2020population}.

A crucial element of this analysis is the correlation time of the environment, which plays an important role in determining the dynamics~\cite{mustonen2008molecular,assaf2013cooperation,wienand2017,cvijovic2015fate}. When the correlation time significantly exceeds the typical timescale of the process in a constant environment (e.g., the typical time for extinction or fixation)  the initial state of the environment is a decisive factor. In this scenario, the probability of fixation, for instance, can be approximated as a weighted average of fixation probabilities in different environmental states. This regime is commonly referred to as the adiabatic or quenched limit~\cite{mustonen2008molecular}. Conversely, when the correlation time is much shorter than the intrinsic demographic timescale, the population samples many different environmental states before fixation occurs, and the initial state of the environment becomes irrelevant. This regime is known as the annealed or white-noise limit, for which a model based on stochastic differential equations with white noise is appropriate. The transition between the quenched and annealed regions is expected to occur when the correlation time of the environment is approximately equal to the intrinsic demographic timescale.

The majority of existing studies have assumed either periodic (seasonal) or Markovian environmental processes, leading to relatively narrow distributions of environmental dwell times. However, numerous works indicate the prevalence of fat-tailed distributions for correlation times associated with various environmental conditions that influence demographic rates, like precipitation, temperatures, etc~\cite{halley1996ecology,ruokolainen2009ecological,fung2018quantifying,vasseur2004color,morales1999viability, fowler2013confounding}. Here we focus on the implications of temporal correlations whose tails are characterized by power-law decay. In view of these results we then explain, quantitatively, what can be expected in other cases featuring fat-tailed statistics of correlation times.

In particular, we uncover a novel intermediate phase that emerges between the quenched and annealed regimes. Within this new phase, the influence of the initial state diminishes, yet the dynamics remain governed by rare (but not too rare) long periods of steady environmental conditions, rather than being dominated by the accumulation of small EN. In what follows we  analyze this intermediate phase and its remarkable properties. Notably, we demonstrate that the width of this phase expands as the tail of the correlation time distribution becomes fatter, ultimately supplanting the annealed phase, when the mean dwelling time diverges. Moreover, the lifespan of an isolated population within this intermediate phase exhibits a nonuniform dependency on the spectral exponent that characterizes the EN, and a power-law scaling with the log of the number of individuals, in stark contrast to the scaling properties observed in Markovian processes.

Our basic example is a well-mixed population of constant size $N$, comprising two strains. At any given time $t$, $n(t)$ individuals belong to strain 1, whose fraction is denoted as $x = n/N$, while the remaining $N-n$ individuals belong to strain 2. In the following we consider the dynamics with an equal initial number of strains, such that $x_0 \equiv x(t=0)=0.5$. In any discrete elementary step, two individuals are selected at random for a duel, the loser dies, the winner produces a single offspring and time is incremented by $1/N$. In the case of an interspecific duel, the probability of species 1 winning is given by $f = e^s/\left(e^s+1\right)$, with the probability of species 2 winning being $1-f$. Here, $s$ represents the selection parameter, implying that species 1 has an advantage when $s>0$. The transition rates are defined as $n\xrightarrow{W_1} n+1$ and $n\xrightarrow{W_2} n-1$, where $W_1 = 2Nx(1-x)f$ and $W_2 = 2Nx(1-x)(1-f)$. On average, as long as $|s| \ll 1$, the fraction $x$ satisfies the equation $\dot{x} = sx(1-x)$.

To model EN, we assume that $s$ varies randomly between two states, $s_{\!-}$ and $s_{\!+}$, such that
\begin{eqnarray}
\label{eq:K}
s(t) =\bar s+\Omega\xi_{\alpha}(t), \;\; \text{with \;$\xi_\alpha(t)\in \{-1,+1\}$,}
\end{eqnarray}
where $\bar s \equiv (s_+ + s_-)/2$ and $\Omega \equiv (s_+ - s_-)/2$.  At stationarity, the mean  $\langle \xi_{\alpha}(t)  \rangle= 0$.

We examine two types of switching time distributions, namely, the waiting time distribution in each environmental state. The simpler case, already discussed in~\cite{Bena06,HL06}, is  a colored {\it symmetric  dichotomous (telegraph) Markov noise}, where the distribution of switching times is exponential, $P_{\rm E}(t) = e^{-t/\tau}/\tau$. We compare this case to the behavior of the system with  Burr (type XII) distributed switching times, 
\begin{equation}
  P_{\rm B}(t)=   ck\lambda\left(\lambda t\right)^{c-1}\left[1+\left(\lambda t\right)^c\right]^{-k-1}. 
\end{equation}
This distribution has  a finite first moment if $\beta>2$, and a power-law tail that decays like $t^ {-\beta}$, where $\beta\equiv kc+1$ is the spectral exponent.  The parameter $\lambda$, which sets the characteristic rate, is related to the mean switching time $\tau$ via  $\lambda= k {\cal B}\left(k-1/c,1/c+1\right)/\tau $, where ${\cal B}(z_1,z_2)=\int_0^1 t^{z_1-1}(1-t)^{z_2-1}dt\,$ is the Beta function. To have a meaningful
comparison between the waiting time distributions, we impose the same mean switching time $\tau$ for both cases. Clearly, as the spectral exponent $\beta$ increases, one expects the Burr system to exhibit a progressively higher resemblance to the exponential system.

Let us consider, e.g., the probability of ultimate fixation $\Pi(x_0)$, where $x_0$ is the initial fraction of species 1. This quantity (also known as exit probability~\cite{redner2001guide}) is key in the theory of population genetics and evolution. When $s$ is fixed, $\Pi$  satisfies~\cite{ewens2004mathematical}
\begin{equation} \label{eq:phi1}
\Pi(x_0,s) = \left(1-e^{-N s x_0}\right)/\left(1-e^{-N s}\right).
\end{equation}
In the annealed limit ($\tau \to 0$, white noise) one expects  $\Pi\to\Pi_{\rm A} \equiv \Pi(\bar s) $, with $\bar{s}$ being the average value of $s$. On the other hand, in the quenched regime ($\tau \to \infty$, adiabatic) the initial state of the environment is decisive and $\Pi\to \Pi_{\rm Q} \equiv \left[\Pi(s_+) + \Pi(s_-)\right]/2 $.

\begin{figure} [h]
\includegraphics[width=8cm]{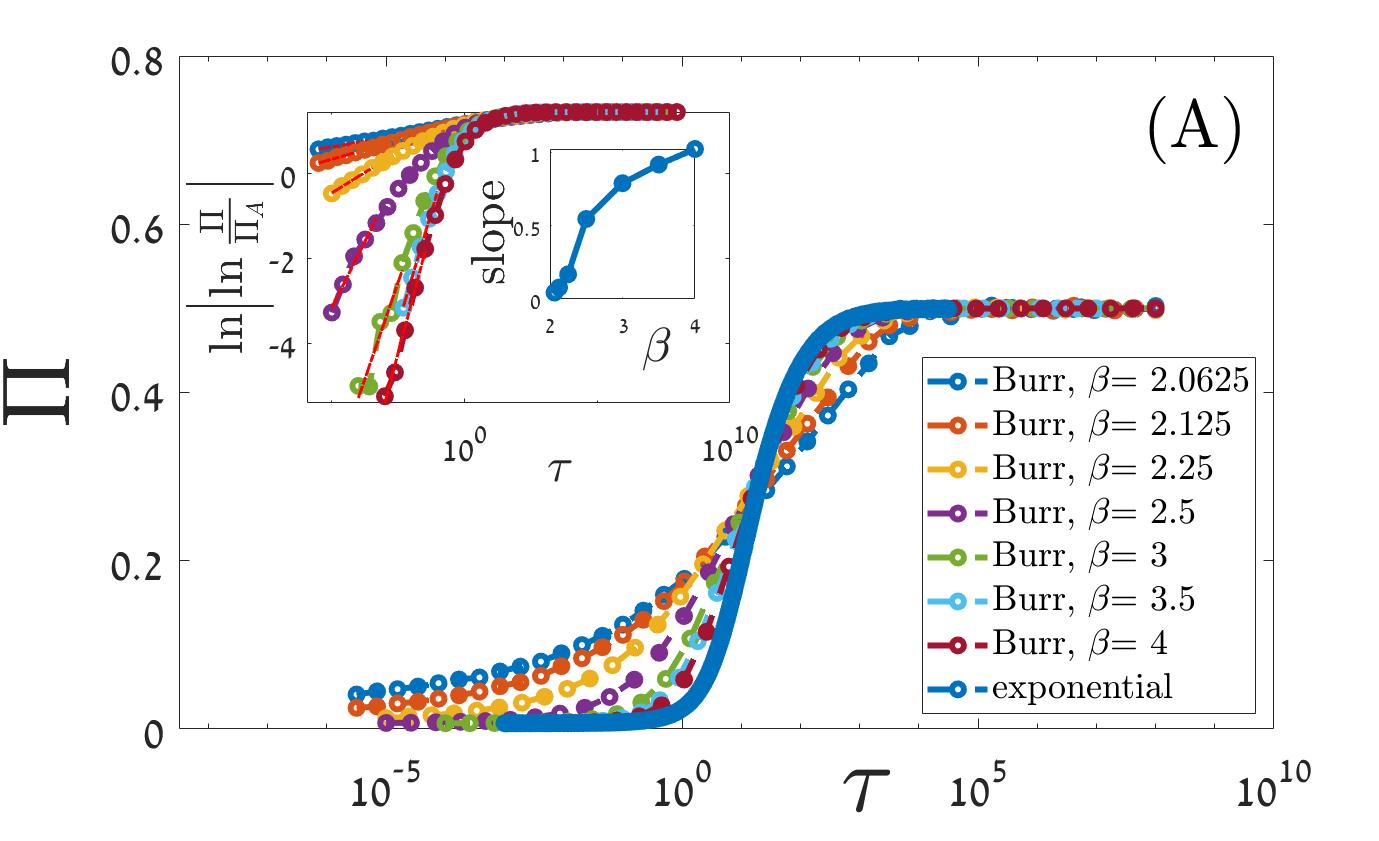}
\includegraphics[width=8cm]{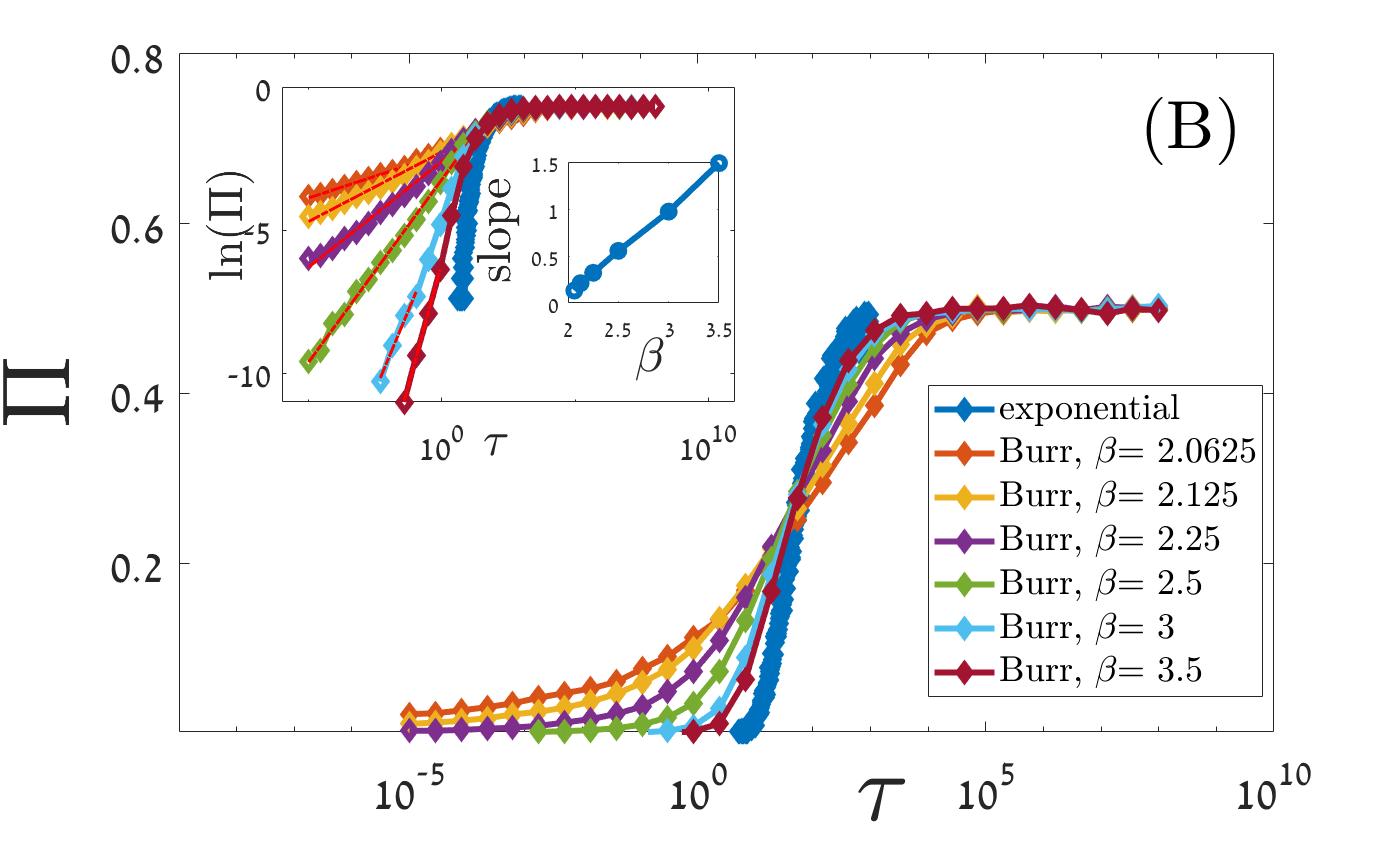}
	\vspace{-4mm}	\caption{\textbf{The intermediate phase.} 
 The fixation probability of the inferior species, $\Pi$, is plotted against the mean dwell time $\tau$.  Selection parameters are $\left(s_+,s_-\right) = \left(0.1,-0.2\right)$. The total number of individuals is $N = 200$ (Panel A) and $N=50000$ (Panel B). The Burr distributions were implemented with $c=1$ and $k=\beta-1$, where  $\beta$ is indicated in the legends. The main panels demonstrate the asymmetric widening of the intermediate phase, while the insets show the convergence to the annealed limit using the scaling suggested in Eq. (\ref{eq10}) for panel (A) (with $\Pi_{\rm A}=0.0069$, according to Eq. (\ref{eq:phi1})), and Eq. (\ref{eq10a}) for Panel (B). Within each inset, we incorporated an additional inset to display the slopes as a function of $\beta$. In panel (A), the slope values noticeably deviate from the predictions, while in panel (B) they do not, see Appendix \ref{ap_hyp} for a detailed explanation.}
\label{fig1}
\end{figure}

The transition between these two limits, the annealed and quenched behaviors, is demonstrated in Fig. \ref{fig1}. Henceforth, we refer to $\Pi$ as the probability of ultimate fixation by the inferior species, and plot $\Pi(x_0=1/2)$ against the mean dwell time $\tau$. As expected, under exponential switching the transition occurs when $\tau$ is close to the typical demographic timescale, which is $1/s$~\cite{blythe2007stochastic}, with a relatively narrow and symmetric transition zone. In contrast, under $P_{\rm B}(t)$, the intermediate phase widens immensely in an asymmetric manner, mainly at the expanse of the annealed regime, which disappears as $\beta \to 2^+$.  The same behavior is demonstrated, in Appendix \ref{ap_Tex},  for the fixation time $T_{\rm fix}$, and in Appendix \ref{ap_con} for a continuously-distributed  $s$, rather than a dichotomous one. We also obtained similar results (not shown) for systems in which $N$ varies over time while $s$ is kept constant.

For the exponential and periodic cases, the convergence of $\Pi(\tau)$ to $\Pi_{\rm A}$ as $\tau \to 0$ was discussed in \cite{taitelbaum2020population}. Here we focus on the fat-tailed (Burr) case, and show that the convergence depends on the ratio between $\Pi_{\rm A}$ and the probability of long dwell times, as explained below.

Intuitively, one realizes that, for a given $\tau$, some realizations (histories) do not include long dwell times while other realizations do. In the former case, the dynamics is annealed-like, while in the latter case the fixation probability is closer to $\Pi_{\rm Q}$. Accordingly, as the chance to pick long dwell times increases (for fixed $\tau$) when the spectral exponent $\beta$ decreases, the intermediate phase widens.

 Now let us attempt to propose a more quantitative analysis. We assume that {\it if}, before the end of the demographic process (fixation), a dwell time $t > T_{\rm Q}$ was picked, the fixation probability is, approximately, $\Pi_{\rm Q}$. When $\tau \to 0$, or when $N$ (and hence $T_{\rm Q}$) are large, the probability of picking such a long dwell time, $q$, is governed by the tail of the distribution, 
 \begin{equation} \label{q}
     q=\int_{T_{\rm Q}}^\infty P(t)dt\approx\left(\lambda T_{\rm Q}\right)^{1-\beta}=\left[\frac{\mu T_{\rm Q}}{\tau}\right]^{1-\beta},
 \end{equation}
where $\mu \equiv \lambda\tau=k {\cal B}\left(k-1/c,1/c+1\right)$. The system reaches fixation through the annealed-like route only when the history of the demographic process is free of these rare events. The typical number of distinct dwell times (or environmental switches) before fixation is $T_{\rm A}/{\tilde \tau}$; here ${\tilde \tau}$ is the mean over the distribution from zero to $T_{\rm Q}$, and it approaches $\tau$ when $T_{\rm Q}/\tau \to \infty$. The probability of a specific history reaching fixation through the annealed route, $P_{\rm A}$, is equal to the chance that, before fixation, a dwell time longer than $T_{\rm Q}$ has never been picked. When the fixation timescales are large and $q \ll 1$ one obtains the stretched exponential dependence, 
\begin{equation} \label{eq9}
P_{\rm A} = (1-q)^{T_{A}/\tau} \sim \exp \left(-\mu^{1-\beta}\frac{T_{\rm A} T_{\rm Q}^{1-\beta}}{\tau^{2-\beta}}\right).
\end{equation}
Thus, given that fixation occurs either via the annealed or quenched routes, the probability of fixation given $\tau$ becomes 
 \begin{equation} \label{eq8}
 \Pi(\tau) = \Pi_{\rm A} P_{\rm A}(\tau) + \Pi_{\rm Q} [1-P_{\rm A}(\tau)].
 \end{equation}
As $\tau \to 0$,  $P_{\rm A} \to 1$, the right term in Eq. (\ref{eq8}) vanishes and one can show that
\begin{equation} \label{eq10}
\ln \left|  \ln (\Pi/\Pi_{\rm A})\right|  \approx (\beta-2) \ln \tau,
\end{equation}
as demonstrated in panel (A) of Fig. \ref{fig1}, (see inset).

In practice, when $N\gg 1$, the fixation probability of an inferior species  in the annealed limit, $\Pi_{\rm A}$, is negligibly small. In that case, for a wide range of $\tau$ values, the right term of Eq. (\ref{eq8}) dominates, yielding
\begin{equation} \label{eq10a}
\ln \Pi(\tau) \approx (\beta-2) \ln \tau,
\end{equation}
and this behavior is demonstrated in panel (B) of Fig. \ref{fig1}, (see inset).

A similar proposition, based on the behavior of the Burr distribution for small arguments, explains the convergence to the quenched limit. The system will be quenched-like when the chance for histories that contain {\it only} short dwell times is negligible. If $t_0$ is such a short enough timescale, the number of independent environments before fixation,  $T_{\rm A}/t_0$, is large, so the chance of picking an annealed-like history decreases exponentially with $T_{\rm A}/t_0$. In particular, for an exponential distribution, as $\tau$ increases the chance to pick $t<t_0 $ is approximately $t_0/\tau$, and therefore the chance for an annealed path to fixation decreases as a power-law in $\tau$,  $\tau^{-T_{\rm A}/t_0}$. Similarly, for Burr distribution, the probability of a short segment is $1-q \approx k(\mu t_0/\tau)^c$, hence one expects again a power-law, $\tau^{-c T_{\rm A}/t_0}$. This convergence is faster than the convergence to the annealed limit and, more importantly, has no singular behavior at $\beta =2$.

Up until now, we have considered a model featuring two absorbing states located at $n=0$ and $n=N$. We will now introduce a reflecting boundary at $n=N$, resulting in the sole absorbing state being the extinction of the focal population at $n=0$. This "ceiling" aligns with the $\theta \to \infty$ variant of the theta-logistic model~\cite{gilpin1973global} and serves as an effective representation of other well-established classical models of single-population dynamics, such as logistic growth. This adjustment enables the exploration of another crucial characteristic of the intermediate phase, namely, the mean time to extinction ($T_{\rm ex}$) of a single population.

 Quite a few results are known for the Markovian versions of this model.  When ${\bar s}<0$, the time to extinction scales like $\ln n_0/\bar{s}$, where $n_0$ is the initial population of the focal species~\cite{yahalom2019comprehensive}. 
If both ${\bar s}>0$ and $s_->0$ (so the population tends to grow even in bad years), extinction occurs only due to demographic noise and $T_{\rm ex}$ grows exponentially with $N$ for $N\gg1$~\cite{kessler2007extinction,assaf2017wkb}. If ${\bar s}>0$ and $s_-<0$, extinction takes place via rare sequences of bad years and $T_{\rm ex}$ follows a power-law in $N$~\cite{yahalom2019comprehensive}. 

\begin{figure} [!t]
		\includegraphics[width=10cm]{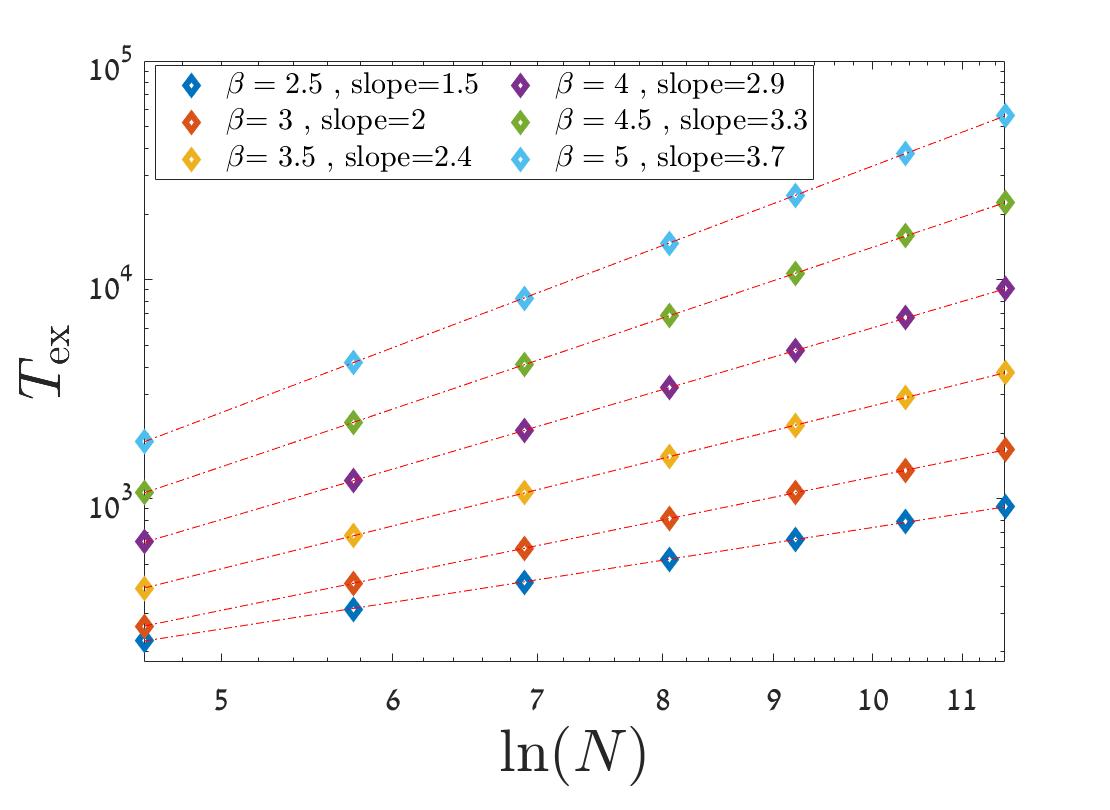}
	\vspace{-4mm}	\caption{\textbf{Extinction times of a stable population.} The time to extinction, denoted as $T_{\rm ex}$, plotted against the natural logarithm of population size, $\ln N$. Both axes employ a logarithmic scale, as indicated by the uneven spacing of tick marks on the x-axis. This display reveals a growth pattern that follows a power-law relationship with $\ln N$, in agreement with the prediction from Eq. (\ref{eq12}). The pink lines represent linear fits to the measured data points (diamonds). The slopes of these fitted lines, as indicated in the legend, are expected to be $\beta-1 = ck$, and the level of agreement between this prediction and the observed slopes is quite remarkable. For all simulations, the dwell time was sampled from a Burr distribution with parameters $c=\lambda=1$ and $k= \beta-1$, while the selection parameter $s$ was drawn from a normal distribution with mean and variance of unity. }
  \label{fig2}
	\end{figure}

When the dwell times are Burr-distributed, a different scaling emerges. 
If $N$ is large, $T_{\rm A} \gg T_{\rm Q}$, and thus the plausible route to extinction is by picking a {\it single} dwell time whose duration is longer than $T_{\rm Q}$. The probability to pick such a long period of stable environmental conditions is $q$, as defined in Eq. (\ref{q}). Therefore when $\beta >2$  the mean time to extinction, $T_{\rm ex}$, scales like,
\begin{equation} \label{eq12}
T_{\rm ex} \sim 1/q \sim (\ln N)^{\beta -1}.
\end{equation}
This novel behavior, a power law in the logarithm of $N$, is demonstrated in Fig. \ref{fig2} for the dichotomous noise case.

Examining extinction times offers us the chance to delve into scenarios where $\beta \le 2$. Within this regime, $\tau$ diverges, and so does the mean time to extinction, whereas the convergence of $T_{\rm ex}$ is not guaranteed.  Yet, we can focus on the average of the {\it logarithm} of the extinction time, which is finite as long as $\beta > 1$. 

\begin{figure} [!t]
		\includegraphics[width=9cm]{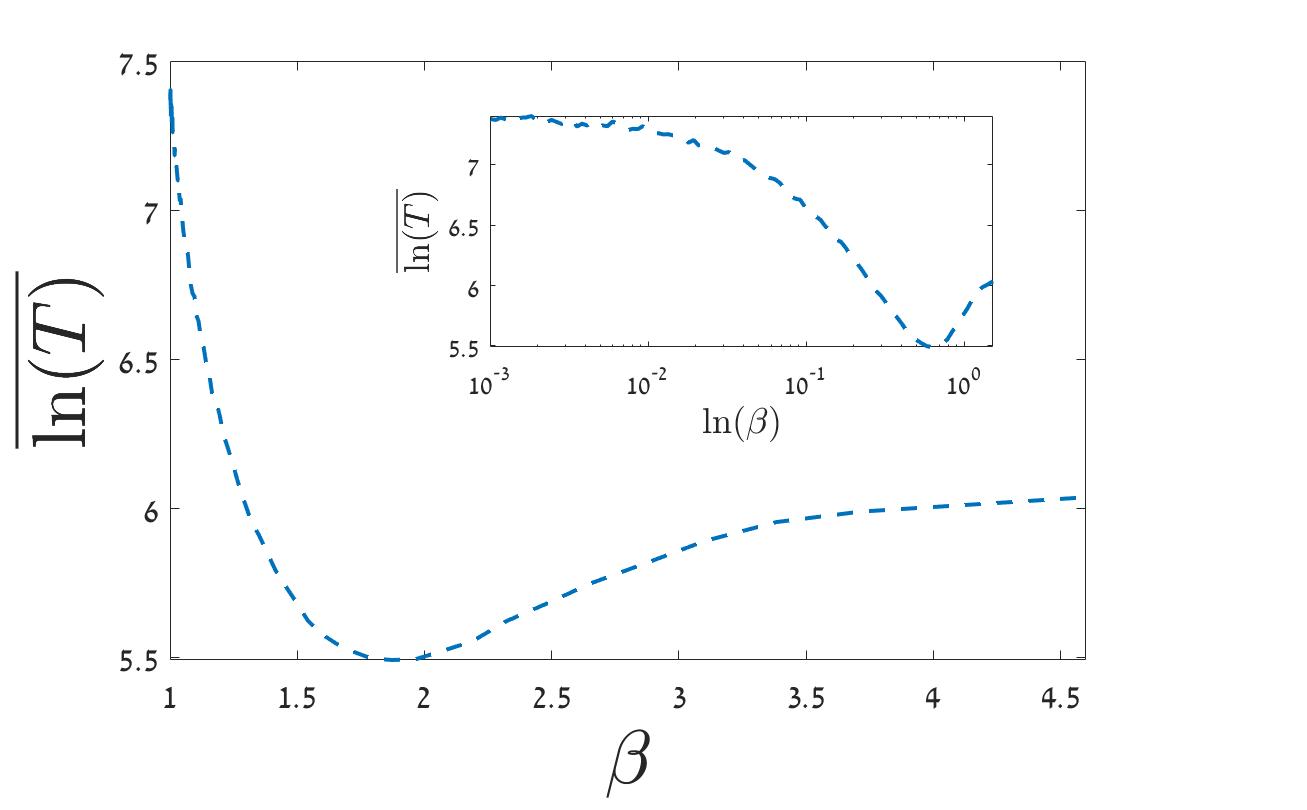}
		\vspace{-5mm}\caption{\textbf{Typical extinction time as a function of $\beta$. } The mean-logrithm of the extinction time is plotted against $\beta$ (main panel). 
  Here, $N=100$, and the Burr distribution parameters are  $c=\lambda=1$, and $k=\beta-1$. The selection parameters are  $\left(s_+,s_-\right) = \left(0.08,-0.04\right)$, such that $\bar{s}>0$. As anticipated (see main text), the point of maximum instability, characterized by the shortest typical extinction time, is observed to be around $\beta=2$. Furthermore, additional simulations (not shown) suggest that as $N$ increases, the minimum point approaches $\beta=2$.  The inset shows the same dataset versus $\ln \beta$, on a semi-logarithmic (in the x-axis) plot, which highlights the plateau near $\beta=1$, where  $\overline{\ln{T_{\rm ex}}}\simeq 7.5$. In the absence of demographic stochasticity, the lifetime of the system diverges as $\beta\to 1$. Yet, when accounting for demographic noise, one finds that in a  constant environment with $s=0.08$, $\overline{\ln{T_{\rm ex}}} \simeq 11$, while for $s=-0.04$,  $\overline{\ln{T_{\rm ex}}} \simeq 4$, whose average is $7.5$.  }\label{fig4}
	\end{figure}

Figure \ref{fig4} reveals a rather unexpected trend: the typical extinction time showcases a non-monotonic behavior with respect to $\beta$, and its minimum at a large $N$ seems to converge towards $\beta=2$. As previously elucidated, when $\beta > 2$, large-$N$ extinction time is dominated by the chance of picking a rare series of bad years with a duration $t > T_{\rm Q}$. As $\beta$ decreases towards $\beta =2$, the probability of such adverse events increases. Conversely, when $\beta < 2$, the system's ``survival'' is primarily governed by the likelihood of picking long periods of favorable environmental conditions. Consequently, in this regime the mean lifespan of the system elongates as $\beta \to 1$.

However, even if a positive survival rate ($s > 0$) is maintained for an arbitrarily long period, each finite population reaches extinction due to demographic stochasticity. The timescale required for these extinction events is exponential in $N$, still, it introduces a cutoff on the divergence of $T_{\rm ex}$ at $\beta \to 1$, as demonstrated in the inset of Fig.~\ref{fig4}.

We have studied a population of constant size comprising two strains, subject to fat-tailed environmental noise (EN), with a power-law  switching-times distribution.
We revealed a novel intermediate phase that emerges between the quenched and annealed regimes. Within this phase, the dynamics is governed by long periods of steady environment, rather than being dominated by the accumulation of small EN. Notably, we showed that this phase's width greatly expands  as the tail of the correlation time distribution becomes wider, ultimately supplanting the annealed phase in the limit $\beta=2$, where the distribution's mean diverges (see Fig.~\ref{fig1}). Furthermore, we found that the lifetime of the system within this intermediate phase scales as a power law in the logarithm of the number of individuals (see Fig.~\ref{fig2}), in contrast to the scaling properties observed in Markovian processes.

Indeed, there are many cases where power laws are not an adequate description of the empirical reality. In these cases, the distribution can be broad but still exhibits an upper cutoff. Here, our results are applicable as long as this cutoff is much longer than the typical timescale of the process in a constant environment. Moreover, our analysis allows to infer the value of the upper cutoff at which the intermediate phase appears. It can also be readily generalized, e.g., to log-normal and stretched-exponential distributions. All in all, considering the intermediate phase described here can shed new light on understanding processes in population genetics, dynamics of communities, and the evolution of species.

The work of I.M. was supported by an Eshkol Fellowship of the Israeli Ministry of Science. M.A. was supported by the Israel Science Foundation Grant No. 531/20. 

\bibliography{ref}
\clearpage

\appendix

\setcounter{figure}{0}
\renewcommand{\thefigure}{S\arabic{figure}}

\section{Fixation Time}\label{ap_Tex}

Here we study 
the mean time to fixation $T_{\rm fix}$  -- the number of elementary time steps until fixation of either species occurs. 
As explained in the main text, when the correlation time is fat-tailed, the likelihood of encountering an annealed process depends on the probability of {\it not} having long dwell times before fixation. This probability diminishes as the typical time to fixation increases and  $\beta$  decreases. Consequently, the annealed phase vanishes as $\beta \to 2$. The quenched limit is reached when the parameter $\tau$ is set such that the probability of selecting {\it only} short segments of a fixed environment becomes exceedingly small.

\begin{figure} [h]
		\includegraphics[width=0.41\textwidth]{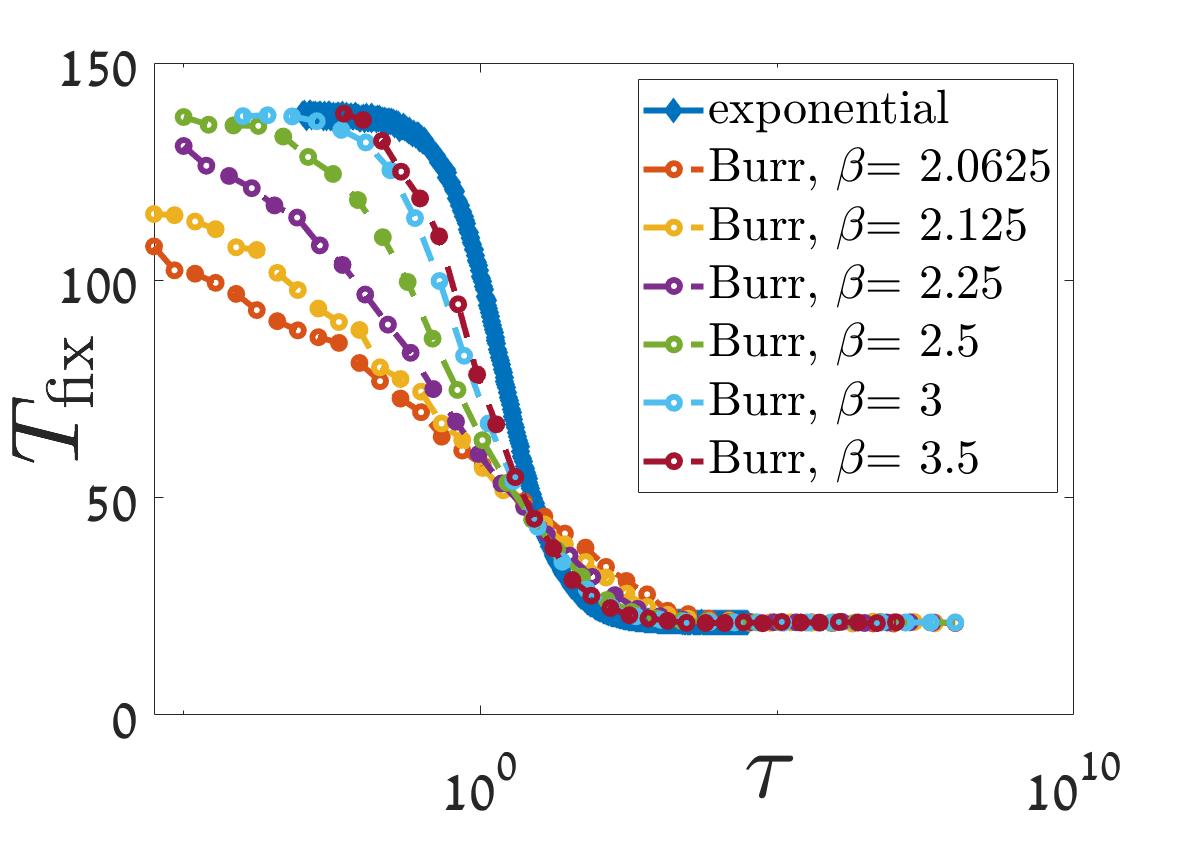}
		\vspace{-4mm}\caption{\textbf{Time to fixation.} 
 Shown is the time to fixation, denoted as $T_{\rm fix}$, versus the mean switching time $\tau$. Here the population size is $N=200$, and $\left(s_+, s_-\right) = \left(0.2, -0.2\right)$.
The plot clearly illustrates the asymmetric broadening of the intermediate phase and the disappearance of the annealed phase as $\beta \to 2$, as shown for $\Pi$ in Fig.~\ref{fig1} of the main text.   }\label{Z33}
  
\end{figure}

In parallel with Fig.~\ref{fig1} of the main text for fixation probabilities, Fig.~\ref{Z33} illustrates this behavior for fixation times. Again, when the correlation function displays fat tails, the intermediate phase expands in an asymmetric fashion and eventually "absorbs" the annealed phase, while the convergence to the quenched phase is notably swift. To accentuate this feature, even in a system with only $N = 200$, we consider the "on average neutral" case where $|s_+| = |s_-|$, as in such a scenario, the time to fixation in the annealed limit scales with $N$, while in the quenched limit, it scales with $\log N$.

\section{Continuous distributions}\label{ap_con}

\begin{figure} [h]
		\includegraphics[width=0.5\textwidth]{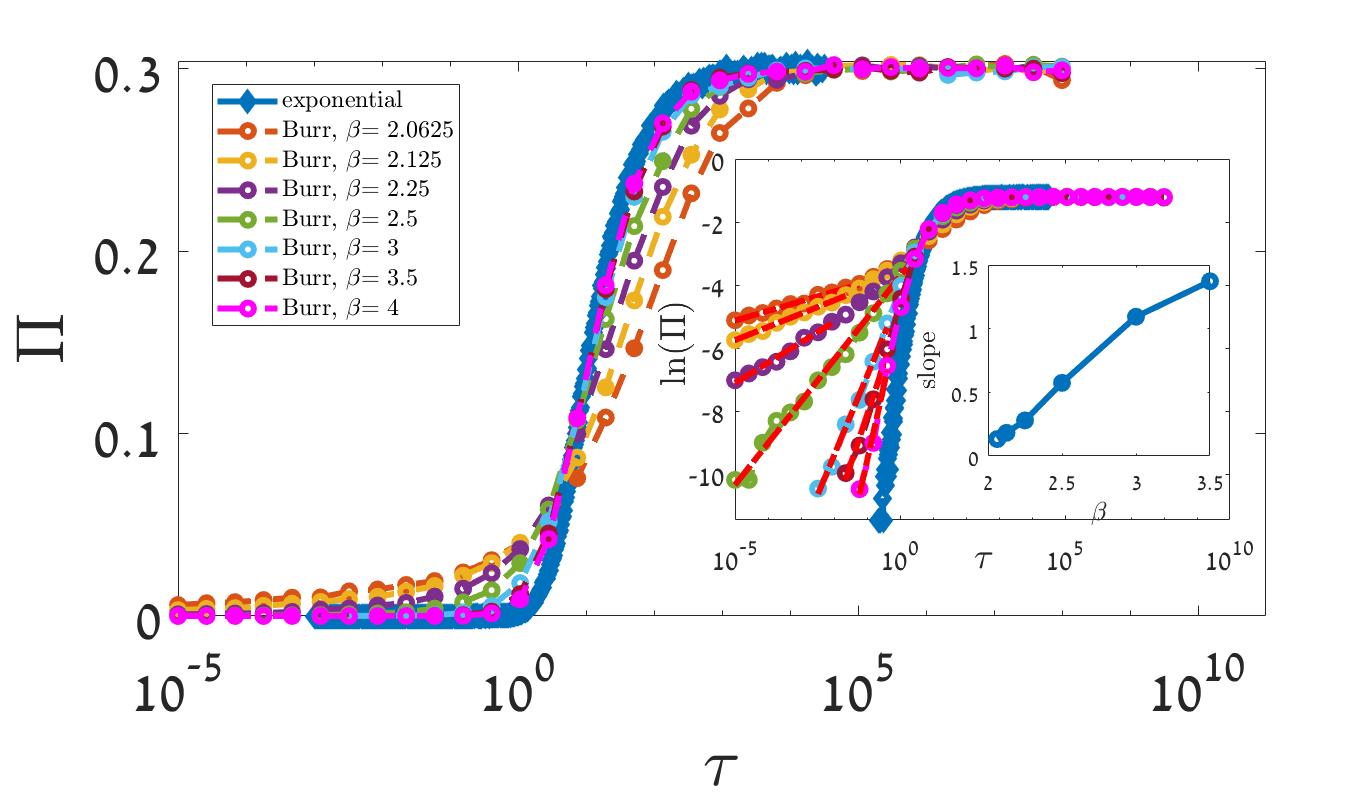}
		\vspace{-4mm}\caption{\textbf{Continuous distributions.} 
  The fixation probability of the inferior species, $\Pi$, is plotted against the mean dwell time $\tau$. Here $N=200$, and the selection parameter $s$ is generated from a uniform distribution between $-0.7$ and $0.3$. In the quenched limit we expect $\Pi$ to follow Eq. (\ref{eq:phi1}), averaged uniformly over all initial conditions, leading to $\Pi_{\rm Q}=0.3$. When $\tau \to 0$, $\Pi$ is tiny (approximately $\exp(-9)$), so we expect the behaviour of of Fig. \ref{fig1} (B), see inset.}\label{Z3}
\end{figure}

Here we present results for the the process outlined in the main text, but with selection values sampled from a continuous uniform  distribution characterized by a specified mean and variance. It is important to note that the mean selection parameter, denoted as $s$, is predominantly negative, although there is a potential for positive selection, as previously discussed in the main text. The outcomes are depicted in Fig.~\ref{Z3}, where we observe a similar trend: an intermediate phase which broadens when the times are sampled from a distribution with fat tails. As the parameter $\beta$ increases, we approach a scenario resembling the exponential time distribution, as depicted in Fig.~\ref{fig1} in the main text.

\section{The disturbance-dictated-outcome hypothesis and its limitations}\label{ap_hyp}
Here we would like to elaborate on the behavior of the slope as function of $\beta$ in the insets of panels A and B of Fig.~\ref{fig1}, and the validity of Eqs.~(\ref{eq10}) and (\ref{eq10a}), for these panels, respectively. In panel (B) the agreement holds for all $\beta$ values, while in panel (A) the slope converge to a limit of 1. This observation aligns with the expected outcome for the exponential distribution, as revealed in previous studies~\cite{taitelbaum2020population, taitelbaum2023evolutionary}. The fundamental intuition behind this is rooted in the fact that
for large values of $\beta$, the effective value of $T_{\rm Q}$ (the lower limit in Eq.~(\ref{q})) becomes $\tau$-dependent, resulting in a change in the exponent. 

The intermediate phase is defined in the main text as the range wherein the likelihood of an improbable event, such as the fixation by an inferior species, hinges on the occurrence probability of a sufficiently large or prolonged disturbance. We quantified the probability of such an event, for example, in Eq. (\ref{q}) of the main text. Our approach operates under the assumption that the system's state preceding the substantial disturbance holds no relevance. In other words, a prolonged period of superiority is posited to result in fixation, irrespective of the system's state before such an event occurs.

As one might anticipate, for a fixed value of $\tau$ the quality of this approximation improves as $\beta$ diminishes, since the smaller $\beta$ is, the time periods are more heterogeneous so the importance of rare events increases. As $\beta$ decreases, a distinct demarcation emerges between minor disturbances, which exert minimal change, and substantial disturbances, where the initial system state holds negligible significance. Conversely, as $\beta$ increases, this distinction becomes less pronounced, and the state of the system immediately preceding a disturbance assumes increasing relevance. Therefore, Eq.~(\ref{eq9}) becomes less accurate as $\beta$ increases. 

To assess the validity of this disturbance-dictated hypothesis, we ran extensive numerical simulations. These  not only tracked the ultimate outcome (fixation of either species) but also monitored $\tau_{\rm max}$, denoting the duration of the longest period of stable environmental conditions before fixation. According to the disturbance-dictated hypothesis, if $\tau_{\rm max} > T_{\rm Q}$ 
, the probability of fixation is $\Pi_{\rm Q}$
. Conversely, when $\tau_{\rm max} < T_{\rm Q}$, the system exhibits behavior akin to an annealed system, and the probability of fixation by the inferior species is $\Pi_{\rm A}$. 

\begin{figure}[h]
		\includegraphics[width=0.65\textwidth]{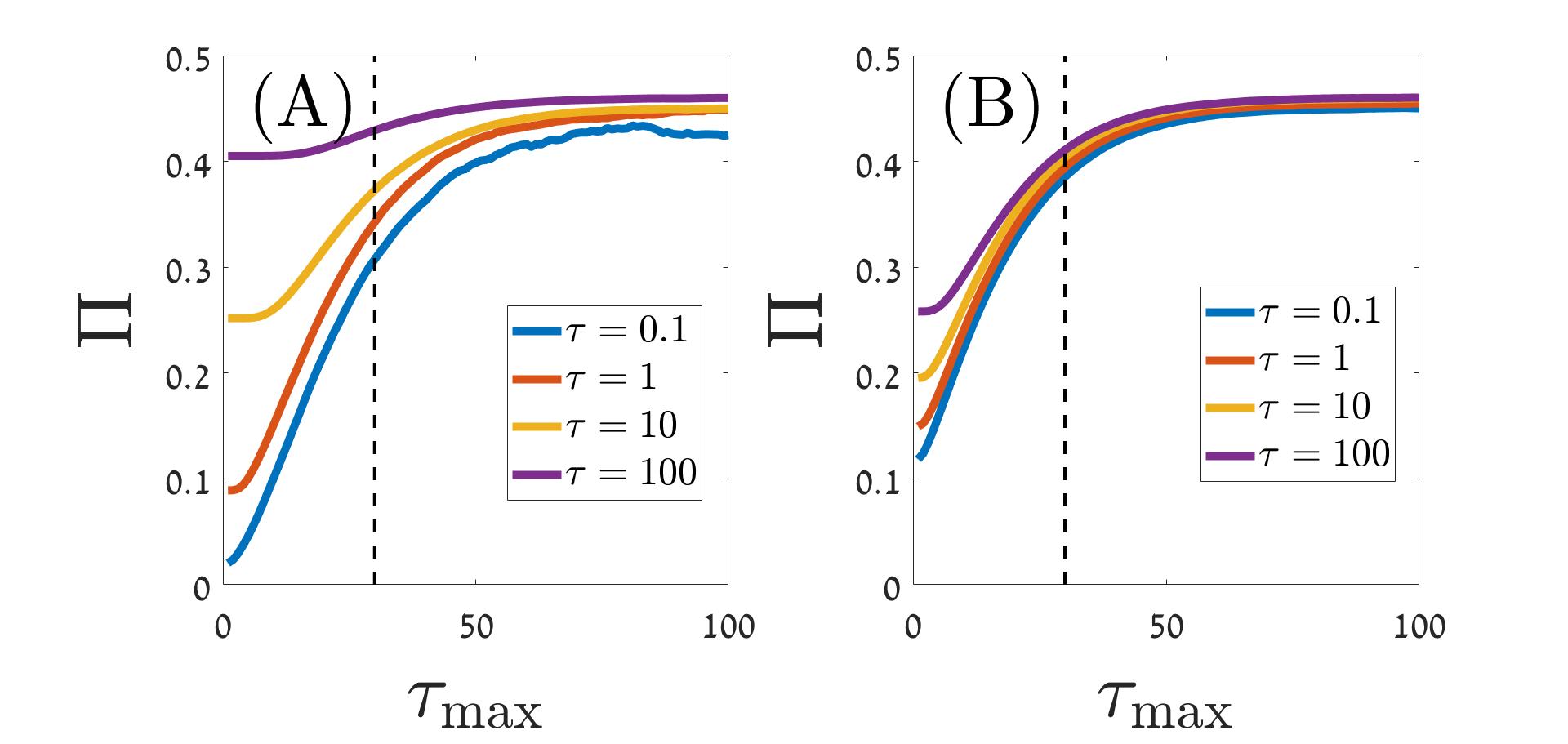}
		\vspace{-4mm}\caption{
  \textbf{The fixation probability versus the maximal dwell time, $\tau_{\rm max}$, in different realizations}.
 Here $N=200$, $\left(s_+,s_-\right) = \left(0.1,-0.2\right)$,  $\beta=3$ in (A) and $\beta=2.005$ in (B), and the legend describes the $\tau$ values. The times to extinction in the quenched limit are $18.5$ (for $s=-0.2$) and $30$ (for $s=0.1$). Therefore, histories that include times longer than $30$ (dotted vertical line) are expected to converge towards the quenched result. 
  }\label{Z1}
	\end{figure}

Figure~\ref{Z1} illustrates the distinctions between the scenarios of large $\beta$ (panel A) to those of $\beta \to 2$ (panel B). Both panels depict the probability of the on-average-inferior species achieving fixation. 
The disturbance-dictated hypothesis implies a clear separation between scenarios where $\tau_{\rm max}>T_{\rm Q}$ and those where $\tau_{\rm max}<T_{\rm Q}$.
In both panels, as $\tau_{\rm max}$ increases, the probability of the inferior species prevailing increases and eventually levels off at the vicinity of $\Pi_{\rm Q}$. However, when $\beta \to 2$, this plateau is almost independent of $\tau$, indicating that once the history includes $\tau_{\rm max}>T_{\rm Q}$, the fixation probability becomes largely independent of other historical factors. Conversely, in the case of larger $\beta$, the spread is considerably wider even when $\tau_{\rm max}>T_{\rm Q}$, implying that the accuracy of the disturbance-dictated hypothesis diminishes. 

\end{document}